\documentclass[conference]{IEEEtran}

\usepackage{wrapfig,enumerate}

  \usepackage[dvips]{graphicx}
  \graphicspath{{./Figures/}}
  \DeclareGraphicsExtensions{.eps}

\usepackage[cmex10]{amsmath}
\usepackage{amsthm}
\usepackage{amssymb}

\usepackage[tight,footnotesize]{subfigure}

\def\E{\mathbb{E}}

\begin{document}

\title{Unlicensed LTE/WiFi Coexistence: \\ Is LBT Inherently Fairer Than CSAT?}

\author{\IEEEauthorblockN{\emph{Cristina Cano, Douglas J. Leith}}
\IEEEauthorblockA{School of Computer Science and Statistics, Trinity College Dublin, Ireland\\
}
}
\maketitle

\begin{abstract}

Ensuring fair co-existence between unlicensed LTE and WiFi networks is currently of major concern to both cellular operators and WiFi providers.  Two main unlicensed LTE approaches currently being discussed, namely Carrier Sense Adaptive Transmission (CSAT) and Listen Before Talk (LBT).  While these mechanisms differ in their compatibility with existing LTE specifications and regulatory compliance in different countries, they also use fundamentally different approaches to access the channel.  Nevertheless, we show in this article that when optimally configured both approaches are capable of providing the same level of fairness to WiFi and that the choice between CSAT and LBT is solely driven by the LTE operator's interests.

\end{abstract}


\IEEEpeerreviewmaketitle

\section{Introduction}\label{sec:introduction}

The 3rd Generation Partnership Project (3GPP) is actively studying the viability of mobile operators using the unlicensed spectrum in order to assist with satisfying increasing mobile traffic demands.  However, a major concern of the 3GPP as well as of WiFi providers and regulatory bodies is the need to ensure fair co-existence with other technologies \cite{flore2014slides,FCC-note,WiFiAllianceStatement}. Given that current technologies in unlicensed bands, such as WiFi \cite{IEEE80211-IEEESTD1999}, rely on carrier-sensing and contention-based access, starvation may occur when they share the channel with a schedule-based technology such as LTE.  

There are, at this stage, two main LTE mechanisms under consideration for ensuring fair co-existence with WiFi. Namely, \emph{Listen Before Talk} (LBT) and \emph{Carrier Sensing and Adaptive Transmission} (CSAT) \cite{rahman2011license,qualcomm2014whitepapers}.   LBT uses carrier sensing and backoff rules in a similar manner to WiFi.   In contrast, CSAT schedules transmissions according to a desired duty-cycle, oblivious to the channel status when a transmission is scheduled to start.    CSAT mainly targets early deployments and the US market, whereas LBT requires changes to the LTE specifications and so is a longer term proposal but is necessary to meet regulations in Europe and Japan.   There have been some preliminary and inconclusive studies, see \cite{jeon2014lte}, on the performance of both approaches and their ability to protect WiFi transmissions. However, CSAT is commonly regarded as being more aggressive and less \emph{fair} than LBT because it does not abide by the same rules as WiFi.

We show in this article that when appropriately configured both mechanisms can provide the same level of fairness to WiFi transmissions.  In particular, we derive the proportional fair rate allocation when using both CSAT/WiFi and LBT/WiFi and establish that in both cases the WiFi airtime is the same.  We confirm this analysis using detailed simulations.  That said, we find that the overhead (in terms of channel time spent in LTE/WiFi collisions) of CSAT is higher than that of LBT.  The proportional fair rate allocation accounts for this inefficiency to LTE and, therefore, it only affects the available LTE airtime and not the WiFi airtime.   Consequently, the choice between using CSAT and LBT is primarily a decision driven by the LTE operator's interests. That is, an LTE operator may select CSAT or LBT based on the LTE throughput, simplicity, operational and management cost, regulatory constraints \emph{etc} but their ability to protect WiFi transmissions is not a design driver.

The remainder of this article is organised as follows. In Section \ref{sec:scenario} we describe the {problem setup}. Then, we derive the throughput model in Section \ref{sec:model} and the proportional fair allocation for CSAT and LBT in Section \ref{sec:prop_fair_allocation}. We show the results in Section \ref{sec:results} and conclude with a discussion on our assumptions and some final remarks in Section \ref{sec:scope} and \ref{sec:conclusions}.


\section{Preliminaries}\label{sec:scenario}

\subsection{LTE Control Overhead \& Subframe Alignment}
Similarly to  \cite{liu2011framework} {and in line with current 3GPP discussions}, we assume that unlicensed LTE control messages are sent via the licensed band.  Since a user needs to receive control information to locate and decode its data within an LTE subframe, it follows that control messages sent over the licensed interface for a given subframe must be aligned with the subframe in the unlicensed band where the data is actually transmitted.  That is, transmitting the control information through the licensed band means that the unlicensed channel transmissions be synchronised to the subframe boundaries in the licensed interface. 

Since the channel access in CSAT is oblivious of the channel status, transmissions can be easily aligned to subframe boundaries in the licensed band, where control information is transmitted. However, given that LBT opportunistically grabs the channel when empty, its channel accesses are not generally aligned with subframe boundaries in the licensed band. As already being considered in the 3GPP, LBT can then transmit a reservation signal until the start of the next subframe boundary in order to make WiFi transmitters refrain from accessing the channel, known as \emph{Load Based Equipment} (LBE). After the reservation signal, the transmission of data can start according to the control information sent in the licensed band.

\subsection{Cost of Heterogeneity}

Note that both CSAT and LBE carry an overhead when co-existing with WiFi.  With CSAT additional WiFi/LTE collisions are generated, and so network throughput is lowered.  With LBE the reservation signal reduces the airtime available for data transmissions which again lowers network throughput.   This overhead can be reduced by increasing the duration of each LTE transmission (as this overhead is a per transmission one). However, increasing the duration of LTE transmissions will tend to increase the delay experienced by WiFi. Thus, the overhead can also be expressed as a trade-off between throughput and delay.   We will return to this trade-off later.
 
%
%

\section{Throughput Model}\label{sec:model}

We begin by developing a throughput model when LTE and WiFi networks share the same wireless channel.   We treat CSAT and LBT/LBE in a unified fashion, to simplify both the model and the later fairness analysis.

\begin{figure}
\centering
\includegraphics[width=0.8\columnwidth]{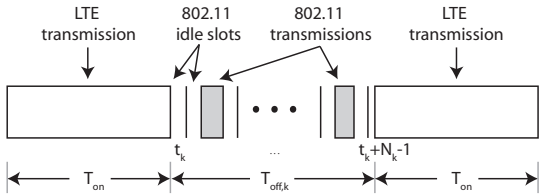}
\caption{Schematic showing LTE/802.11 transmission timing.}\label{fig:slots2}
\end{figure}

We assume that multiple LTE networks/operators use different channels that do not interfere with one another, which is in line with current 3GPP discussions.   When modelling throughput and analysing fairness between LTE and WiFi it is therefore sufficient to consider a single LTE network coexisting with one or more WiFi networks.  Let $n$ denote the number of WiFi stations sharing a channel with an unlicensed LTE network.   Let $T_k$, $k=1,2,\dots$ denote the times when LTE transmissions start.  Each LTE transmission is of duration $T_{\rm on}$, so the LTE silent/off interval between transmissions $k$ and $k+1$ is of duration $T_{{\rm off},k}:=T_{k+1}-T_k-T_{\rm on}$, see Figure \ref{fig:slots2}.   We assume that random variables $T_{{\rm off},k}$, $k=1,2,\dots$ are i.i.d with mean $\bar{T}_{{\rm off}}:=\E[T_{\rm off}]$.  We also assume that 802.11 stations sense the channel as being busy during an LTE transmission and so no new 802.11 transmission start during an LTE $T_{\rm on}$ period. 
Note that there may be an LTE/802.11 collision at the start of a $T_{\rm on}$ period when LTE starts transmitting while an 802.11 transmission is already in progress.

\subsection{802.11 MAC Slots}
During the $T_{{\rm off},k}$ period when LTE is silent following the end of a $T_{\rm on}$ period, the 802.11 stations perform their usual CSMA/CA random access.   The 802.11 MAC partitions time into MAC slots which may be either an idle slot, of duration $\sigma$, or a busy slot, of duration $T_{\rm b}+\rm{DIFS}$ (for simplicity, we assume that both successful 802.11 transmissions and collisions between 802.11 transmissions are of the same duration), where $T_{\rm b}$ denotes the time to transmit a packet, including the frame transmission ($T_{\rm fra}$), {\rm SIFS} and ACK ($T_{\rm ack}$).   We index these 802.11 MAC slots during the $T_{{\rm off},k}$ period by $t_k,t_{k}+1,\dots,t_{k}+N_k-1$, see Figure \ref{fig:slots2}.   

Note that at the end of the $T_{{\rm off},k}$ period there will generally be a partial 802.11 MAC slot, since the end of the $T_{{\rm off},k}$ period is \emph{not} aligned with the 802.11 MAC slot boundaries, but $t_{k}+N_k$ indexes the last full MAC slot.  The number of MAC slots $N_k:=t_{k+1}-t_k$ in the $T_{{\rm off},k}$ period is a random variable.

\subsection{802.11 Events}
Let $Z_{t,j}$ be a random variable which takes the value $1$ when 802.11 station $j$ transmits in MAC slot $t$.  We assume that the $Z_{t,j}$, $t=1,2,\dots$ are i.i.d, $Z_{t,j}\sim Z_{j}$ and let $\tau_j:=Prob(Z_j=1)$.  We also assume that the $Z_{t,j}$, $j=1,\dots,n$ are independent.   

Let $X_{t}$ be a random variable which takes the value $1$ when MAC slot $t$ is busy ($Z_{t,j}=1$ for at least one $j\in\{1,\dots,n\}$), and $0$ otherwise.  The $X_{t}$, $t=1,2,\dots$ are i.i.d, $X_{t}\sim X$, with $p_{\rm e}:=Prob(X=0)=\prod_{i=1}^n(1-\tau_i)$.   
Since the $X_{t}$, $t=1,2,\dots$ are i.i.d and the $T_{{\rm off},k}$, $k=1,2,\dots$ are also i.i.d. the number $N_k$ of MAC slots in the $T_{{\rm off},k}$ periods $k=1,2,\dots$ are i.i.d, $N_k\sim N$.    The duration of MAC slot $t$ is $M_t:=\sigma+X_{t}(T_{\rm b}+\rm{DIFS}-\sigma)$.  The $M_t$, $t=1,2,\dots$ are i.i.d, $M_t\sim M$ with $\E[M]=\sigma p_{\rm e}+(T_{\rm b}+\rm{DIFS})(1-p_{\rm e})$.

Let $Y_{t,j}$ be a random variable which takes the value $1$ when there is a successful (non-colliding) transmission by 802.11 station $j$ in MAC slot $t$, and $0$ otherwise.    The $Y_{t,j}$, $t=1,2,\cdots$ are i.i.d, $Y_{t,j} \sim Y_j$, with $p_{succ,j}:=Prob(Y_{j}=0)=\frac{\tau_j}{1-\tau_j}p_{\rm e}$.   
The number of successful transmissions in the $T_{{\rm off},k}$ period is $W_{k,j}:=\sum_{t=t_k}^{t_{k}+N_k-1} Y_{t,j}$ and the mean rate in bit/s of 802.11 station $j$ is $s_{{\rm wifi},j}:=\lim_{K\rightarrow\infty}\frac{\sum_{k=1}^K W_{k,j}}{\sum_{k=1}^K T_{\rm on}+T_{{\rm off},k}} D_j$, where $D_j$ is the number of data bits communicated by station $j$ in a successful transmission.

\subsection{802.11 Throughput}
The $W_{k,j}$, $k=1,2,\cdots$ are i.i.d, $W_{k,j}\sim W_j$, and the $T_{{\rm off},k}$, $k=1,2,\cdots$ are also i.i.d, $T_{{\rm off},k}\sim T_{{\rm off}}$  (but note that $W_{k,j}$ and $T_{{\rm off},k}$ are not independent since the number of successful transmissions depends on the duration of the $k$'th off period).   The $W_{k,j}$, $T_{{\rm off},k}$, $k=1,2,\cdots$ define a renewal-reward process and it follows that  $s_j =\frac{\E[W_{j}]}{T_{\rm on}+\E[T_{{\rm off}}]} D_j$.
We have that $\E[W_{j}] = \E[Y_{j}]\E[N]$ since $Y_{j}$ and $N$ are independent, and $\E[Y_{j}]=p_{succ,j}$.  It remains to determine $\E[N]$ (the average number of full 802.11 MAC slots in an LTE off period).

Let $\hat{T}_{{\rm off},k}=\sum_{t=t_k}^{t_{k}+N_k-1}M_t$.  That is, $\hat{T}_{{\rm off},k} \le T_{{\rm off},k}$ is the duration of that part of the $T_{{\rm off},k}$ period occupied by full 802.11 MAC slots \emph{i.e.} excluding any partial MAC slot at the end of the period when LTE lacks carrier sensing.  It follows that $\E[N] = \frac{\E[\hat{T}_{{\rm off}}]}{\E[M]}$ since the $M$ is independent of $N$.   Hence,
\begin{align}
s_{{\rm wifi},j} = \frac{p_{succ,j}}{\sigma p_{\rm e}+(T_{\rm b}+\rm{DIFS})(1-p_{\rm e})}\frac{\E[\hat{T}_{{\rm off}}]}{T_{\rm on}+\bar{T}_{{\rm off}}} D_j.
\end{align}
Observe that $s_j:=\frac{p_{succ,j}}{\sigma p_{\rm e}+(T_{\rm b}+\rm{DIFS})(1-p_{\rm e})}D_j$ is just the usual expression for the throughput of an 802.11 station \cite{checco2011proportional}, but that this is now scaled by $\frac{\E[\hat{T}_{{\rm off}}]}{T_{\rm on}+\bar{T}_{{\rm off}}}$.

\subsection{$\E[\hat{T}_{{\rm off}}]$}
\subsubsection{CSAT}
CSAT does not make use of of carrier sensing and so an LTE transmission may start part way through an 802.11 MAC slot.   In this case we might approximate $\E[\hat{T}_{{\rm off}}]$ by $\E[T_{{\rm off}}]$, and we can expect this approximation to be accurate when $\E[T_{{\rm off}}]$ is sufficiently large that any partial 802.11 MAC slots can be neglected.  However, when $\E[T_{{\rm off}}]$ is smaller it is necessary to use a more accurate approximation for $\E[\hat{T}_{{\rm off}}]$.   We adopt the following.    When the start times $T_k$, $k=1,2,\dots$ of the LTE transmissions satisfy the lack of anticipation property, \emph{e.g.} when the spacing $T_{k+1}-T_k$ is drawn from an exponential distribution \cite{baccelli2006role}, then the LTE transmissions satisfy the PASTA property.   
That is, the probability that the start of an LTE on period coincides with an 802.11 transmission is $p_{{\rm LTE}} = \frac{p_{\rm s} T_{\rm b}+ p_{\rm c} T_{\rm fra} }{\E[M]}$.   Assuming that on average the start of an LTE transmission that collides with an 802.11 transmission occurs half-way through the 802.11 transmission, then

\small
\begin{align}
\E[\hat{T}_{{\rm off}}]&=\E[T_{{\rm off}}](1-p_{{\rm LTE}}) + (\E[T_{{\rm off}}] - \frac{T_{\rm fra}}{2})p_{{\rm LTE}}\\
&=\bar{T}_{{\rm off}} - \frac{T_{\rm fra}}{2}p_{{\rm LTE}}.
\end{align}
\normalsize

\subsubsection{LBT/LBE}
With LBE, the start of an LTE on period is aligned with an 802.11 MAC slot boundary since LBE uses carrier sense to ensure this.   Therefore there are no partial MAC slots and $\E[\hat{T}_{{\rm off}}]=\E[T_{{\rm off}}]$.   Also, the  probability that the start of an LTE on period coincides with an 802.11 transmission is just $p_{{\rm LTE}} = 1-p_{\rm e}$, that is, the probability of having at least one 802.11 station transmitting in a given MAC slot.

\subsection{{\rm LTE} Throughput}
Let $r$ denote the LTE transmit rate in bits/s.  When the start time of an LTE transmission does not coincide with an 802.11 transmission then the error-free LTE transmission is of duration $T_{\rm on}$ \emph{i.e.} $rT_{\rm on}$ LTE bits are transmitted.   When the LTE start time coincides with an 802.11 transmission then we assume that the first part of the LTE transmission is lost.  The precise behaviour differs for CSAT and LBE, as follows.

\subsubsection{CSAT}
On average the start of an LTE transmission that collides with an 802.11 transmission occurs half-way through the 802.11 transmission, and so on average the first $T_{\rm fra}/2$ seconds of the LTE transmission are lost.  Assuming that partial overlap of an LTE subframe with an 802.11 transmission leads to loss of the whole subframe, then $r(T_{\rm on}-\lceil\frac{T_{\rm fra}}{2T_{{\rm LTE}}}\rceil T_{{\rm LTE}})$ LTE bits are transmitted, where $T_{{\rm LTE}}$ denotes the duration of an LTE subframe.  It follows that the LTE throughput when using CSAT is:
\begin{align}
s_{{\rm LTE}} &=r\frac{T_{\rm on}(1-p_{{\rm LTE}}) + (T_{\rm on}-\lceil\frac{T_{\rm fra}}{2T_{{\rm LTE}}}\rceil T_{{\rm LTE}})p_{{\rm LTE}}} {T_{\rm on}+\bar{T}_{{\rm off}}}\\
&=r\frac{T_{\rm on}-\lceil\frac{T_{\rm fra}}{2T_{{\rm LTE}}}\rceil T_{{\rm LTE}}p_{{\rm LTE}}} {T_{\rm on}+\bar{T}_{{\rm off}}}.
\end{align}

\subsubsection{LBT/LBE}
Since LTE transmissions are aligned with 802.11 MAC slots the duration of an LTE/802.11 collision is simply $T_{\rm fra}$.  Additionally, since the LTE network has to transmit a reservation signal until the next subframe boundary of average duration $T_{\rm res}=T_{{\rm LTE}}/2$, useful LTE data transmission only occurs during $T_{\rm on}-T_{\rm res}$ and the number of bits LTE transmits at each channel access when it suffers from a collision with 802.11 is: $r(T_{\rm on}-\max(T_{\rm res}, \lceil\frac{T_{\rm fra}}{T_{{\rm LTE}}}\rceil T_{{\rm LTE}}))$.  It follows that the LTE throughput when using LBE is:

\small
\begin{align}
s_{{\rm LTE}} &=r\frac{T_{\rm on}-(\max(T_{\rm res}, \lceil\frac{T_{\rm fra}}{T_{{\rm LTE}}}\rceil T_{{\rm LTE}}))p_{{\rm LTE}} - T_{\rm res}(1-p_{{\rm LTE}})}{T_{\rm on}+\bar{T}_{{\rm off}}}.
\end{align}
\normalsize

\section{Proportional Fair WiFi/LTE Allocation}\label{sec:prop_fair_allocation}

We now use the throughput model from the previous section to derive the proportional fair rate allocation when LTE and WiFi share a channel and (i) LTE uses CSAT and (ii) LTE uses LBT/LBE.

\subsection{CSAT}
Let $c_1:=\frac{T_{\rm fra}}{2}p_{{\rm LTE}}$, $c_2:=\lceil\frac{T_{\rm fra}}{2T_{{\rm LTE}}}\rceil T_{{\rm LTE}}p_{{\rm LTE}}$, $z:=\bar{T}_{{\rm off}} - c_1$ and $\tilde{z}:=\log z$.   Also $\tilde{s}_{{\rm wifi},j} :=\log {s}_{{\rm wifi},j}$ and $\tilde{s}_{{\rm LTE}}:=\log {s}_{{\rm LTE}}$.  Then,
\begin{align}
\tilde{s}_{{\rm wifi},j} &= \log s_j \frac{\bar{T}_{{\rm off}} - c_1}{T_{\rm on}+\bar{T}_{{\rm off}}} 
= \log s_j +\tilde{z}-\log(T_{\rm on}+c_1+e^{\tilde{z}}),\notag
\end{align}
and
\small
\begin{align}
\tilde{s}_{{\rm LTE}}&=\log r\frac{T_{\rm on}-c_2}{T_{\rm on}+\bar{T}_{{\rm off}}} 
=\log (r(T_{\rm on}-c_2)) -\log(T_{\rm on}+c_1+e^{\tilde{z}}).\notag
\end{align}
\normalsize
It can be verified (by inspection of the second derivative) that $\log(T_{\rm on}+c_1+e^{\tilde{z}})$ is convex in $\tilde{z}$ when $T_{\rm on}+c_1\ge 0$.  Hence, putting the network constraints in standard form,

\small
\begin{align}
\tilde{s}_{{\rm wifi},j} -\log s_j -\tilde{z}+\log(T_{\rm on}+c_1+e^{\tilde{z}}) &\le 0,\ j=1,\dots,n\\
\tilde{s}_{{\rm LTE}}-\log q +\log(T_{\rm on}+c_1+e^{\tilde{z}})&\le 0,
\end{align}
\normalsize
where $q:=r(T_{\rm on}-c_2)$, it can be seen that they are convex in decision variables $\tilde{s}_{{\rm wifi},j}$, $\tilde{s}_{{\rm LTE}}$ and $\tilde{z}$.

The proportional fair rate allocation for CSAT is the solution to the following utility optimisation,

\small
\begin{align*}
&\max_{\tilde{s}_{{\rm wifi},j},\tilde{s}_{{\rm LTE}},\tilde{z}} \tilde{s}_{{\rm LTE}}+\sum_{j=1}^n\tilde{s}_{{\rm wifi},j}\\
s.t.\quad&\tilde{s}_{{\rm wifi},j} -\log s_j -\tilde{z}+\log(T_{\rm on}+c_1+e^{\tilde{z}}) \le 0,\ j=1,\dots,n\\
&\tilde{s}_{{\rm LTE}}-\log q +\log(T_{\rm on}+c_1+e^{\tilde{z}})\le 0.
\end{align*}
\normalsize
The optmisation is convex and satisfies the Slater condition, hence strong duality holds.  The Lagrangian is,
\begin{align*}
L=&-\tilde{s}_{{\rm LTE}}-\sum_{j=1}^n\tilde{s}_{{\rm wifi},j} \\
&+\theta(\tilde{s}_{{\rm LTE}}-\log q +\log(T_{\rm on}+c_1+e^{\tilde{z}}))\\
&+ \sum_{j=1}^n\lambda_j(\tilde{s}_{{\rm wifi},j} -\log s_j -\tilde{z}+\log(T_{\rm on}+c_1+e^{\tilde{z}})).
\end{align*}
The main KKT conditions are
\begin{align}
&-1+\theta = 0, \quad
-1+\lambda_j = 0\,\ j=1,\dots,n\\
&(\theta + \sum_{j=1}^n\lambda_j)\frac{e^{\tilde{z}}}{T_{\rm on}+c_1+e^{\tilde{z}}}-\sum_{j=1}^n\lambda_j=0.
\end{align}
Thus, at an optimum $\theta=1$, $\lambda_j=1$, $j=1,\dots,n$ and
\begin{align}
\frac{e^{\tilde{z}}}{T_{\rm on}+c_1+e^{\tilde{z}}}=\frac{n}{n+1} \label{eq:toff}.
\end{align}
It can be verified (by inspection of the first derivative) that the LHS is monotonically increasing in $\tilde{z}$ and so a unique solution $\tilde{z}$ exists satisfying (\ref{eq:toff}).   Letting $\tilde{z}^*$ denote this solution, the proportional fair $\bar{T}_{{\rm off}}$ value is given by 
$\bar{T}_{{\rm off}}^* = e^{\tilde{z}^*}+\frac{T_{\rm fra}}{2}p_{{\rm LTE}}$.
The airtime fraction available for full 802.11 MAC slots is
\begin{align}
\frac{\bar{T}_{{\rm off}}^* -c_1}{T_{\rm on}+\bar{T}_{{\rm off}}^*} 
&= \frac{e^{\tilde{z}^*}}{T_{\rm on}+c_1+e^{\tilde{z}^*}} 
= \frac{n}{n+1}\label{eq:p1}
\end{align}
and the LTE airtime fraction is
\begin{align}
1-\frac{\bar{T}_{{\rm off}}^* -c_1}{T_{\rm on}+\bar{T}_{{\rm off}}^*} 
&=\frac{T_{\rm on}+c_1}{T_{\rm on}+\bar{T}_{{\rm off}}^*} 
=\frac{1}{n+1}.\label{eq:p2}
\end{align}

Observe that $c_1$ is the average airtime expended on LTE/WiFi collisions at the start of an LTE transmission.   Letting $T_{on}+c_1$ denote the effective LTE airtime (including both the time spent actually transmitting and the time spent on collisions) and $(T_{off}-c_1)/n$ denote the effective airtime of a WiFi station, then (\ref{eq:p1})-(\ref{eq:p2}) tell us that the proportional fair rate allocation equalises these effective airtimes.   That is, the airtime allocated to the LTE network is the same as that allocated to a WiFi station.   This seems quite intuitive, the most interesting point being that the proportional fair allocation assigns the airtime cost of an LTE/WiFi collision to the LTE network.   Note also that the extension of this analysis to allow multiple users in the LTE network is straightforward, and in this case the airtime allocated to an LTE client is the same as that allocated to a WiFi station (again accounting for the LTE/WiFi airtime cost within the LTE airtime).

\subsection{LBT/LBE}
For LBE the analysis is almost the same as for CSAT but with $c_1:=0$ and $c_2:=\max(T_{\rm res}, \lceil\frac{T_{\rm fra}}{T_{{\rm LTE}}}\rceil T_{{\rm LTE}})p_{{\rm LTE}} - T_{\rm res}(1-p_{{\rm LTE}})$.  We obtain that the proportional fair allocation for LBE has LTE airtime satisfying:

\begin{align}
\frac{T_{\rm on}}{T_{\rm on}+\bar{T}_{{\rm off}}^*} =\frac{1}{n+1}.\label{eq:p3}
\end{align}

\subsection{Discussion}
Bearing in mind that $c_1=0$ for LBE, it can be seen that (\ref{eq:p3}) is identical to (\ref{eq:p2}).   Recall, however, that the LBE $T_{\rm on}$ time includes the time spent transmitting a reservation signal until the next LTE subframe boundary is reached.   Hence, in (\ref{eq:p3}) the cost of the reservation signal is included in the LTE airtime.   That is, for both CSAT and LBE the cost of heterogeneity (the time spent in LTE/WiFi collisions for CSAT, and the time spent on the reservation signal for LBE) is accounted for in the LTE airtime.   Therefore, it follows that the co-existence mechanism used does not have an impact on WiFi airtime when a proportional fair rate allocation is used.

\section{Example}\label{sec:results}

We illustrate the proportional fair WiFi/LTE result for CSAT and LBE in an example network.  As in \cite{liu2011framework}, for simplicity we suppose that all stations (both WiFi and LTE) use the same physical layer and, in particular, a 64-QAM modulation and a 5/6 coding scheme which provides a $135$ Mbps data rate when using a $40$ MHz channel at the $5$ GHz ISM band as defined in IEEE 802.11ac \cite{IEEE80211ac}.  $T_{\rm fra}$ and $T_{\rm ack}$ are calculated as \cite{IEEE80211ac}:

\begin{equation}
T_{\rm fra}=T_{\rm plcp}+ \left\lceil \frac{L_{\rm s} + n_{\rm agg}(L_{\rm del} + L_{\rm mac-h} + D) + L_{\rm t}}{n_{\rm sym}} \right\rceil T_{\rm sym},
\end{equation}

\begin{equation}
T_{\rm ack}=T_{\rm plcp}+\left\lceil \frac{L_{\rm s} + L_{\rm ack} + L_{\rm t}}{n_{\rm sym}} \right\rceil T_{\rm sym},
\end{equation}
where $n_{\rm sym}$ is the number of bits per OFDM symbol, $T_{\rm sym}$ is the symbol duration, $n_{\rm agg}$ is the number of packets aggregated in a WLAN transmission and the rest of parameters are specified in Table \ref{tbl:param}. We consider the transmission probability of a WiFi station to be fixed and equal to $\tau=1/16$.  For the LTE network we consider that the Control Format Indicator (CFI) is equal to 0 (recall that we assume that the control information is sent through the licensed interface).

\begin{figure*}[t!] 
\centering
\subfigure[$n=1,T_{\rm on}=10$ms]{\includegraphics[width=0.65\columnwidth]{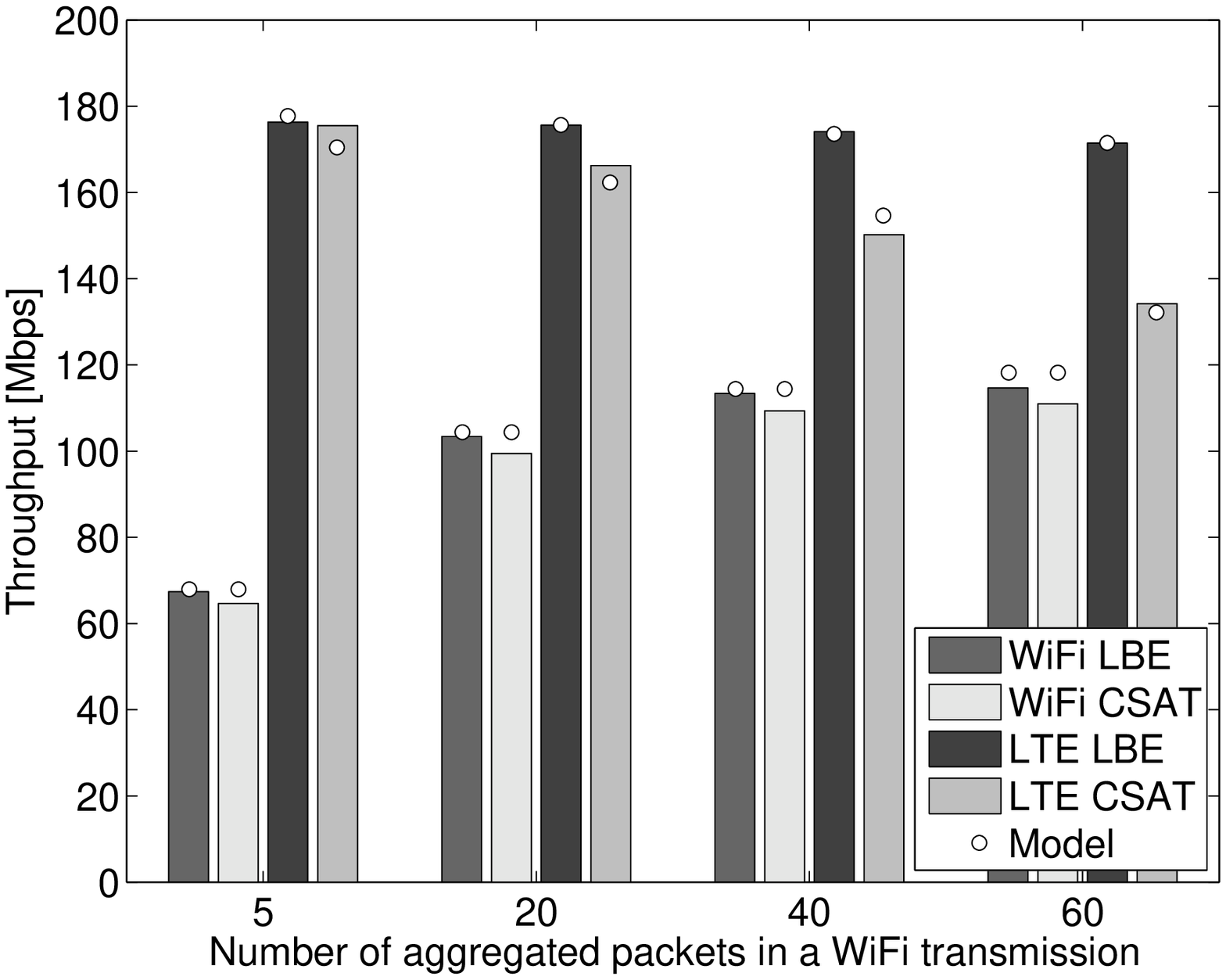}}
\subfigure[$n=3,T_{\rm on}=10$ms]{\includegraphics[width=0.65\columnwidth]{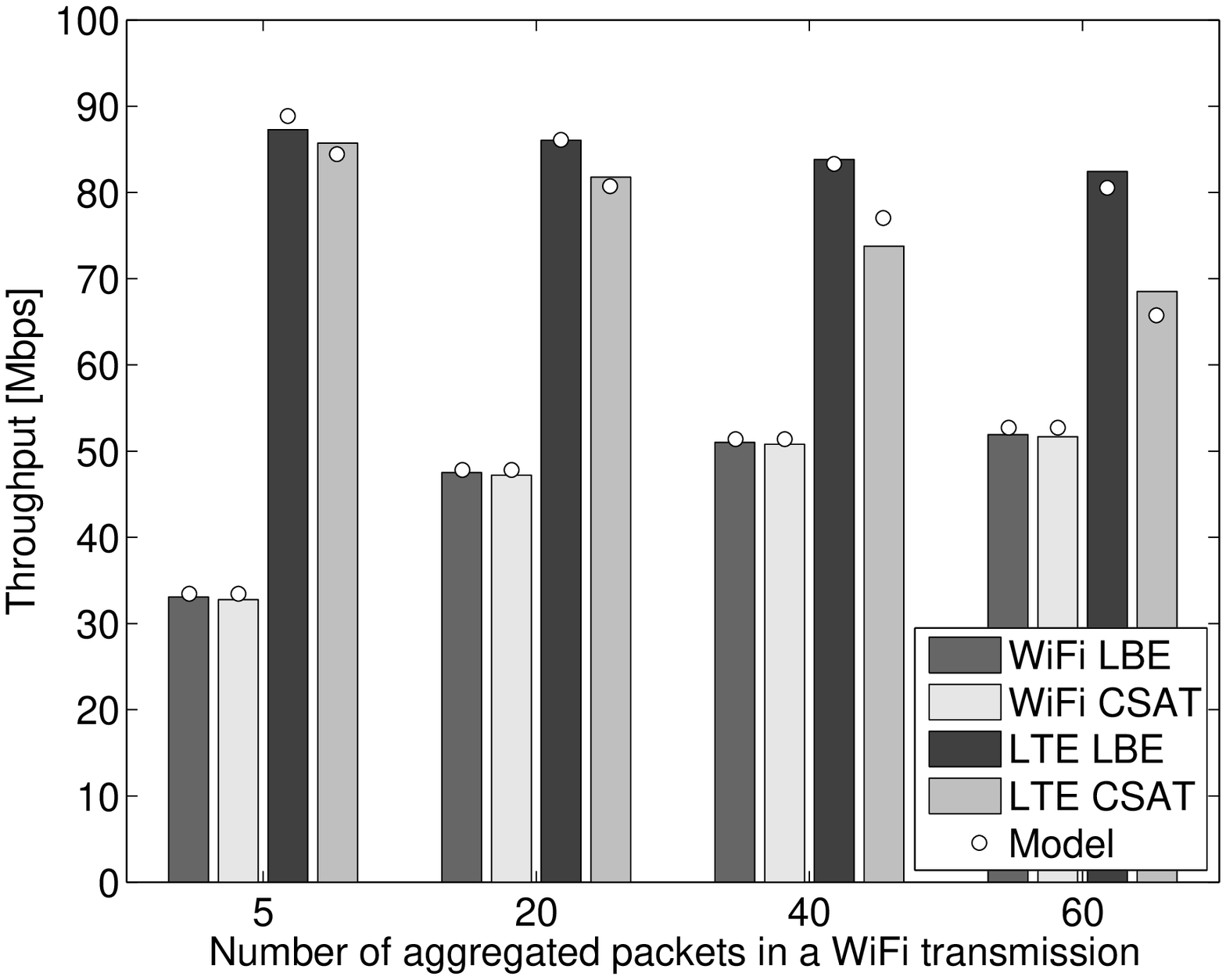}}
\subfigure[$n=9,T_{\rm on}=10$ms]{\includegraphics[width=0.65\columnwidth]{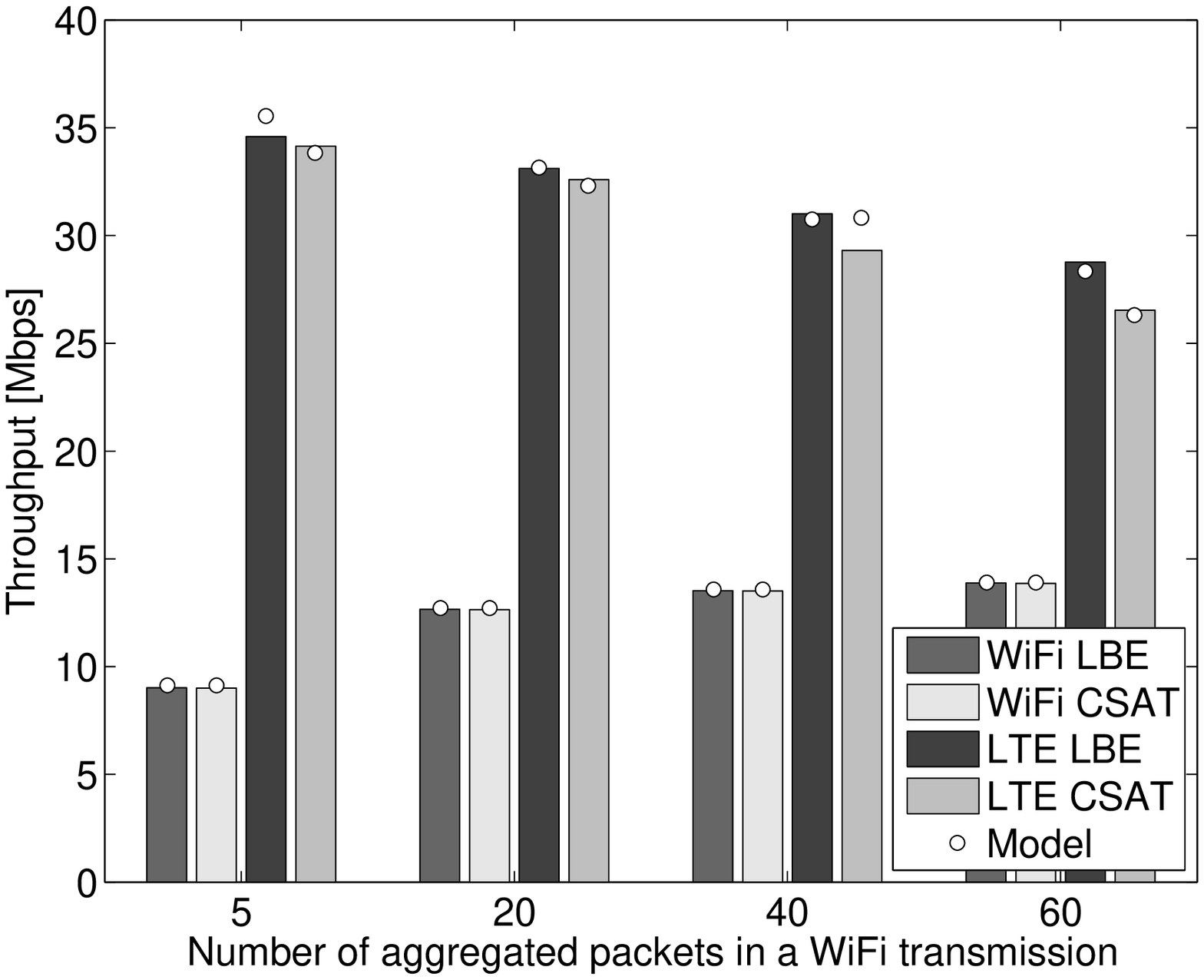}}\\
\subfigure[$n=1,T_{\rm on}=50$ms]{\includegraphics[width=0.65\columnwidth]{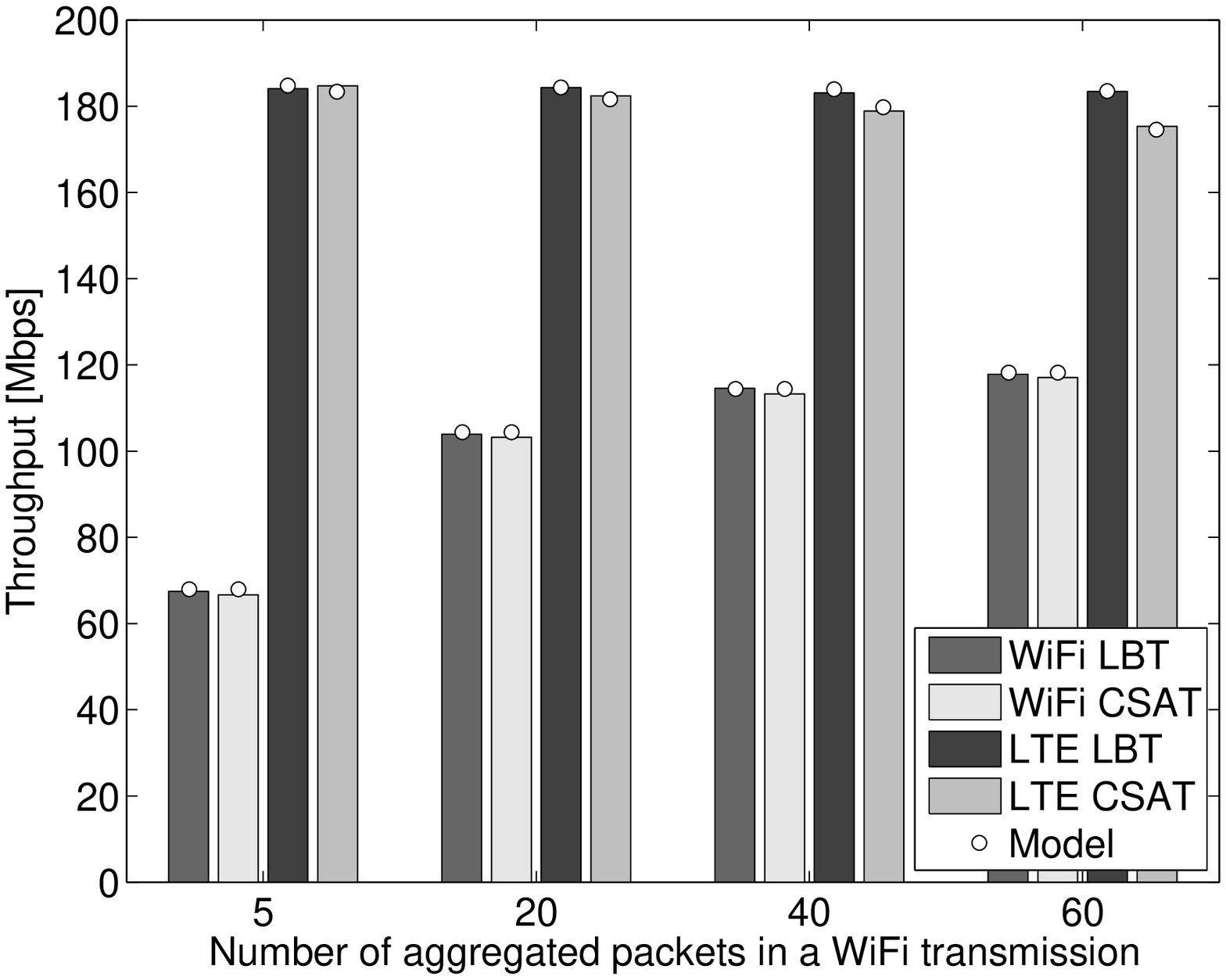}}
\subfigure[$n=3,T_{\rm on}=50$ms]{\includegraphics[width=0.65\columnwidth]{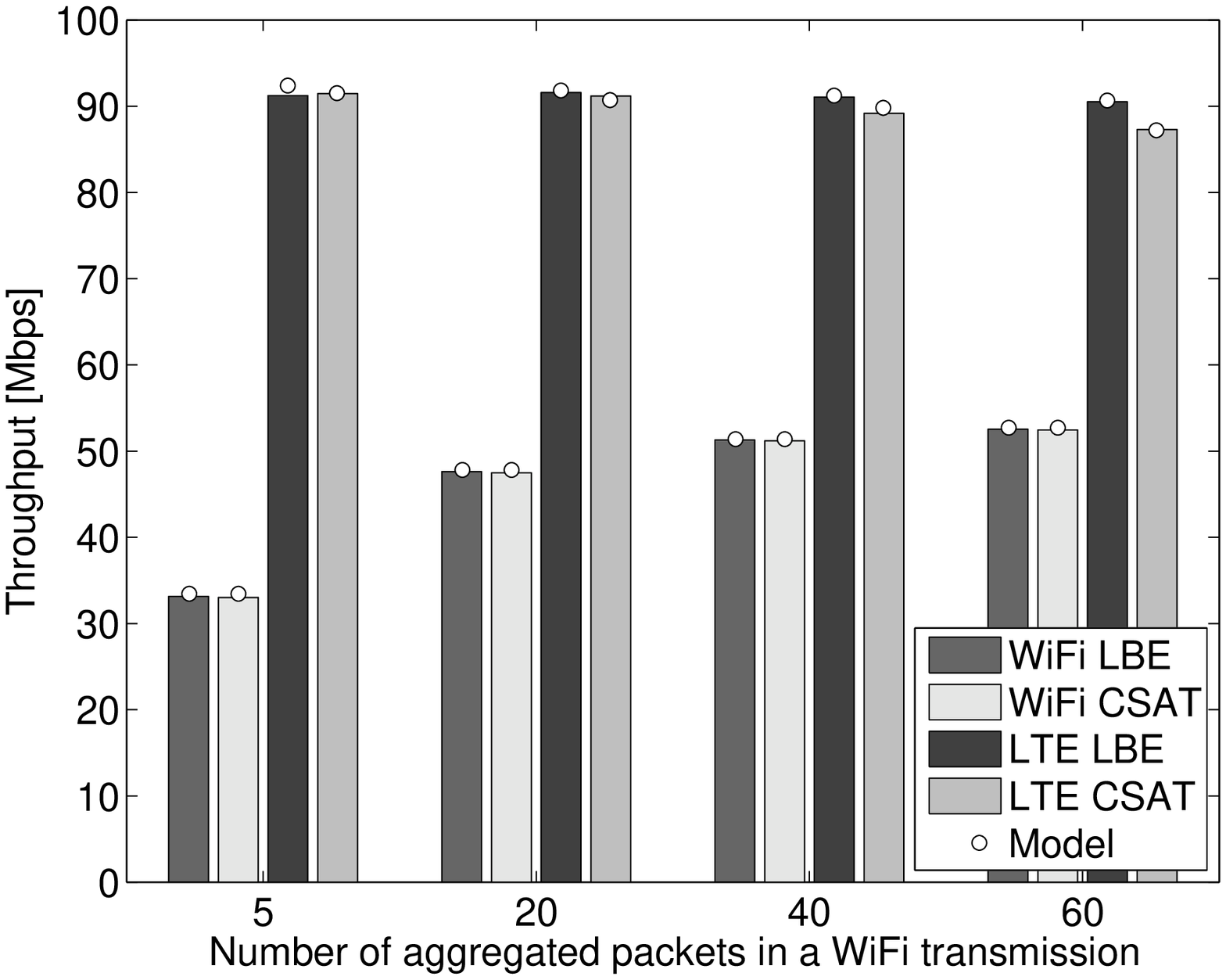}}
\subfigure[$n=9,T_{\rm on}=50$ms]{\includegraphics[width=0.65\columnwidth]{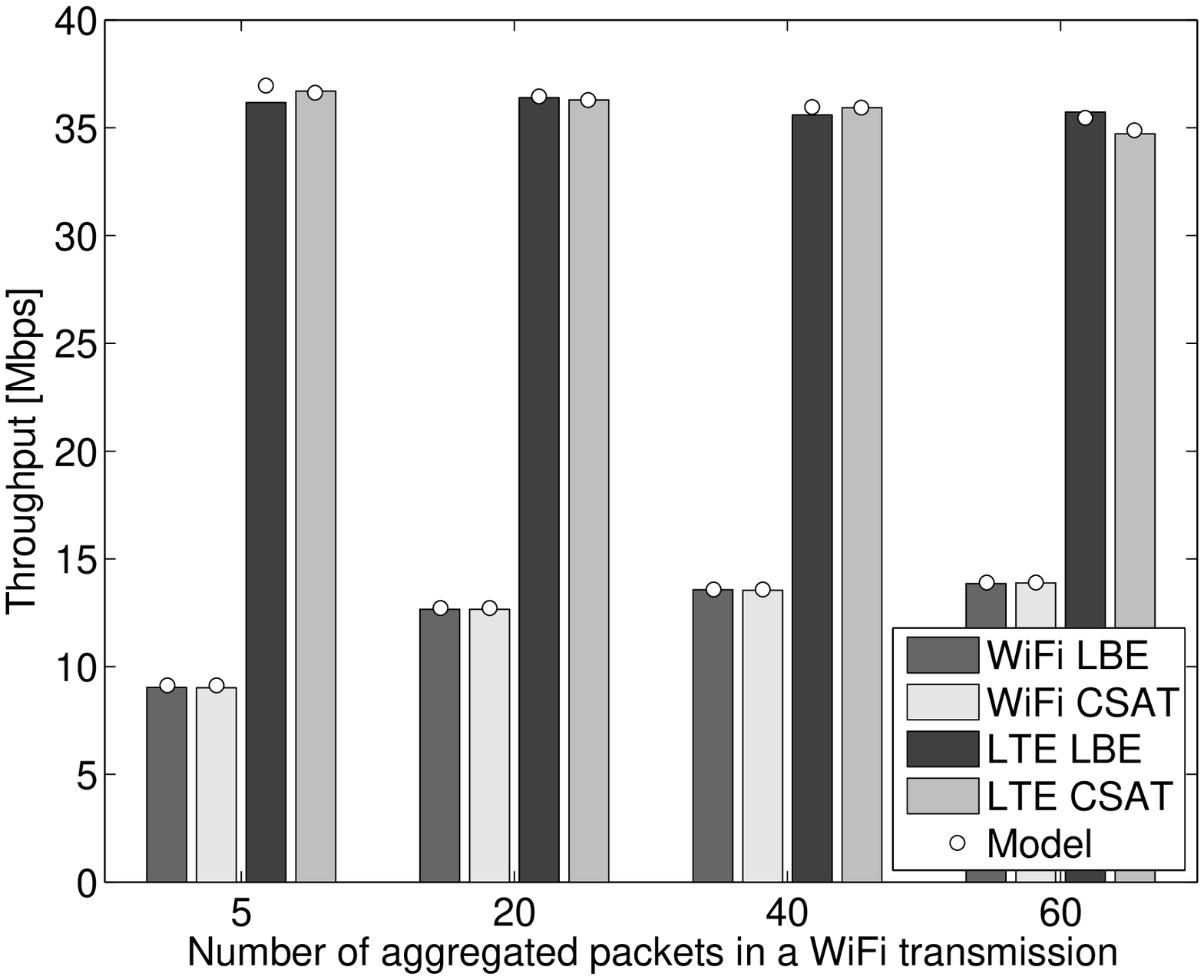}}
\caption{Proportional fair throughput allocation for different configurations of $n$ and $T_{\rm on}$ while varying $n_{\rm agg}$ (effectively changing the packet size of WiFi transmissions). Simulation results are averages of $100$ simulation runs with $50$ s time horizon.}
\label{fig:throughput_prop_fair}
\end{figure*}

\begin{table}[tb]
\centering
\caption{Parameters IEEE 802.11ac \cite{IEEE80211ac}}\label{tbl:param}
\begin{tabular}{c|c} 
Slot Duration ($\sigma$) & $9$~$\mu$s \\ \hline
DIFS & $34$~$\mu$s\\ \hline
SIFS & $16$~$\mu$s\\ \hline
PLCP Preamble+Headers Duration ($T_{\rm plcp}$) & 40~$\mu$s  \\ \hline
PLCP Service Field ($L_{\rm s}$) & 16 bits \\ \hline
MPDU Delimiter Field ($L_{\rm del}$) & 32 bits  \\ \hline
MAC Header ($L_{\rm mac-h}$) & 288 bits  \\ \hline
Tail Bits ($L_{\rm t}$) & 6 bits  \\ \hline
ACK Length ($L_{\rm ack}$) & 256 bits  \\ \hline
Payload ($D$) & $12000$ bits \\ \hline
\end{tabular}
\end{table}

Figure \ref{fig:throughput_prop_fair} shows the WiFi and LTE proportional fair throughputs when using CSAT and LBE.   Results are shown both for detailed packet-level simulations and for the throughput model presented in Section \ref{sec:model}.   These show the impact of varying $T_{\rm on}$, $n$ and the number of aggregated packets in a WiFi transmission (effectively changing the packet size).  It can be seen that the WiFi throughput is essentially the same when using either CSAT and LBE for all configurations.   In contrast, however, the LTE throughput varies depending on the co-existence mechanism used and the network conditions.   For example, we can observe a considerable decrease in throughput when CSAT is used for $T_{\rm on}=10$ms and larger WiFi packet sizes (see Figures \ref{fig:throughput_prop_fair}a-c). The reason for this is the increased collision probability of CSAT compared to LBE.  As already pointed out, the cost of heterogeneity can be reduced by increasing the duration of the LTE transmissions and so it can be seen that both schemes provide similar LTE throughput for $T_{\rm on}=50$ms (Figures \ref{fig:throughput_prop_fair}d-f).

\begin{figure}[ht!] 
\centering
\includegraphics[width=0.70\columnwidth]{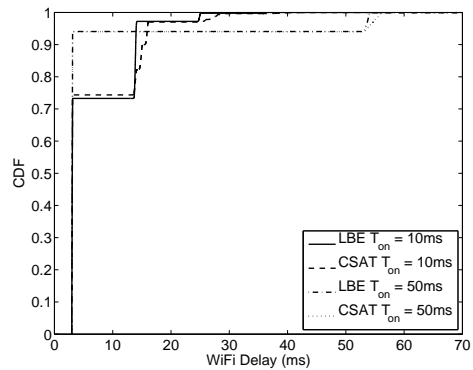}
\caption{CDF of delay for WiFi nodes with $n=1$, $n_{\rm agg}=64$ packets and for different $T_{\rm on}$ values. Results obtained from $100$ simulation runs with $50$ s time horizon.}
\label{fig:cdf_delay}
\end{figure}

Although increasing the duration of the LTE transmissions improves LTE throughput and reduces the cost of heterogeneity, it also causes the delay of WiFi to increase.  This is because WiFi stations defer their transmissions while LTE transmissions are ongoing.   We evaluate the distribution of the MAC access delay of WiFi packets when LTE uses CSAT and LBE in order to assess this throughput-delay trade-off.  Figure \ref{fig:cdf_delay} shows the CDF of the WiFi MAC access delay when $n=1$, $n_{\rm agg}=64$ packets and for $T_{\rm on}=10$ms and $T_{\rm on}=50$ms. It can be seen that for a given value of $T_{\rm on}$, the distribution of the WiFi delay is similar for both CSAT and LBE. We can also see that increasing $T_{\rm on}$ causes longer delays for a fraction of the WiFi packets (namely, those whose transmisison has been deferred while an LTE transmission is in progress). Interestingly, increasing $T_{\rm on}$ while maintaining the proportional fair configuration also causes the LTE network to access the channel less often, that is $T_{\rm off}$ also increases correspondingly. The consequence of this is that a higher percentage of the WiFi packets can access the channel during $T_{\rm off}$, experiencing short delays and so the \emph{mean} WiFi packet delay actually falls as the LTE $T_{\rm on}$ increases.   However, a small fraction of WiFi packets experience long delays.  For example, for $T_{\rm on}=10$ms it can be seen from Figure \ref{fig:cdf_delay} that around $73\%$ of the WiFi transmissions observe short delays, while for $T_{\rm on}=50$ms, this percentage increases to $\sim 94\%$. 

\section{Scope}\label{sec:scope}
In our analysis we have made a number of assumptions, many of which can be fairly readily relaxed.  
%
%
\emph{(i) Lossy channel}: Extension of our model to include channel losses is straightforward.  Namely, by {reducing} the WiFi success probability with a packet loss probability and similarly the unlicensed LTE success probability.  \emph{(ii) Unsaturated stations}: Our model assumes that the WiFi and unlicensed LTE stations are saturated, \emph{i.e.} there is always a packet buffered for transmission. Extension to unsaturated stations can be achieved by adding offered load constraints to the utility-fair optimisation in Section \ref{sec:prop_fair_allocation}.   \emph{(iii) Multiple Channels/Channel Bonding}:  An unlicensed LTE network may in general transmit on multiple WiFi channels.   Similarly, future WiFi networks are expected to make use of channel bonding to transmit across multiple 20MHz channels.   However, provided the WiFi networks occupy disjoint channels, we can solve the unlicensed LTE allocation problem separately for each set of channels using the model in Section \ref{sec:model}.  That is, although we focus on a single channel here, the generalisation to multiple channels is immediate.    \emph{(iv) Perfect WiFi Carrier Sensing of LTE}: Although it has been reported that detection of LTE transmissions by the WiFi carrier sensing mechanism is not always effective \cite{jian2015coexistence}, we assume in this work that WiFi is able to reliably detect LTE transmissions and thus, defer its channel access attempts when the medium is busy due to LTE transmissions. Extensions to the analysis presented here to include some probability of detection by WiFi are straightforward. However, it is still not clear under which conditions WiFi carrier sense fails to detect LTE transmissions. It is important to point out that mechanisms such as the CTS-to-self \cite{sadek2015extending} can be used in LTE to ensure WiFi reliably detects LTE transmissions.

We have also made a number of assumptions which are less easy to relax.  \emph{(i) Unlicensed LTE Channel Widths}:  Although optional channel widths smaller than 20MHz are also being considered by the 3GPP, in this work we consider that both LTE and WiFi use 20MHz channels. The extension to smaller LTE channel widths is not straightforward at present as it is not yet clear the level of interference that each technology will cause to one another when using heterogeneous and partially overlapping channel widths \cite{jian2015coexistence}.  \emph{(ii) Capture}: Our model assumes that concurrent transmissions result in a collision and the inability of the receiver (either unlicensed LTE or WiFi) to decode the message.   The main difficulty with including capture effects in our analysis (where some receivers may successfully decode a colliding transmission) lies in specifying a suitable physical layer model and so we leave this for future work. \emph{(iii) Hidden Terminals}:  Perhaps the most significant omission from our analysis is hidden terminals.     The basic difficulties here arise from the fact that  hidden terminals can start transmitting even when a transmission by another station has already been in progress for some time and that the times hidden terminals attempt transmission are coupled to the dynamics of the transmissions they overhear.  We therefore leave consideration of WiFi/unlicensed LTE allocation with hidden terminals to future work.   It is perhaps also worth noting here that the prevalence of severe hidden terminals in real network deployments presently remains unclear.  While it is relatively easy to construct hidden terminal configurations in the lab that exhibit gross unfairness, it may be that such configurations are less common in practice.

\section{Concluding Remarks}\label{sec:conclusions}

In this work we evaluate the main two co-existence mechanisms under consideration in the 3GPP to provide fairness to WiFi in the presence of an unlicensed LTE network, namely CSAT and LBE. We derive the proportional fair rate allocation to illustrate that when appropriately configured both mechanisms can provide the same level of fairness to WiFi. Therefore, the selection of the co-existence mechanism is primarily driven by the LTE operator's interests, which might include LTE throughput, simplicity, operational and management costs as well as strategical decisions on market targets.

Our analysis also shows that for sufficiently long LTE transmission times, the LTE throughput with both CSAT and LBE is almost identical. However, for shorter LTE transmission times, we find that CSAT provides lower LTE throughput than LBE due to the higher LTE/WiFi collision probability of the CSAT approach. We also evaluate the impact of the LTE transmission time on the distribution of the WiFi MAC access delay. While shorter LTE transmission time decreases the tail of the WiFi delay distribution, the percentage of packets that suffer from long delays increases. The effects of these sporadic and long delays on higher layers, especially on TCP dynamics, requires further understanding.

\section{Acknowledgments}
This work was supported by Science Foundation Ireland under Grant No. 11/PI/1177.

\bibliographystyle{IEEEtran}

\bibliography{references}

\end{document}